\documentclass[%
reprint,
superscriptaddress,
 amsmath,amssymb,
 aps,
 prl,
]{revtex4-1}

\usepackage{graphicx}
\usepackage{dcolumn}
\usepackage{bm}

\begin{document}


\title{
$Z_Q$ Topological Invariants for Polyacetylene, Kagome and Pyrochlore lattices
}%

\author{Y. Hatsugai}\email{y.hatsugai@gmail.com}
\affiliation{Institute of Physics, University of Tsukuba, 1-1-1 Tennodai, Tsukuba, Ibaraki 305-8571, JAPAN}
\author{I. Maruyama}%
\affiliation{%
Graduate School of Engineering Science,
Osaka University, Toyonaka, Osaka 560-8531, JAPAN 
}%


\date{\today}

\begin{abstract}
Adiabatic $Z_Q$ invariants by quantized Berry phases
are defined
for gapped electronic systems in $d$-dimensions ($Q=d+1$).
This series includes 
Polyacetylene, Kagome and Pyrochlore lattice respectively for 
$d=1,2$ and $3$.
The invariants are quantum $Q$-multimer order parameters to characterize 
the topological phase transitions by the multimerization.
 This fractional quantization is protected 
by the global $Z_Q$ equivalence.
As for the chiral symmetric case, a topological form of the 
$Z_2$-invariant is explicitly given as well. 
%
\end{abstract}

\pacs{Valid PACS appear here}
\maketitle


\newcommand{\mynote}[1]{}
\newcommand{\Tr}{{\rm Tr} \, }
\newcommand{\mmat}[4]
{
\left(
\begin{array}{cc}
#1 & #2 \\
#3 & #4 
\end{array}
\right)
}

{\em Introduction}: 
Although the symmetry breaking has been quite successful for a description 
of phases of matter, 
recent development in condensed matter physics reveals
that there are classes of physically important phases
where the symmetry breaking does not play any fundamental roles.
Quantum Hall (QH) effects are historical  examples.
Although  there are many apparently different QH states,
none of them are characterized by the symmetric properties,
where topological quantities, such as, fractional statistics of
quasi-particles \cite{Arovas84}, topological degeneracy\cite{Wen89}
 and the Chern  numbers\cite{Thouless82,NTW}
as the topological invariants are fundamentally important. 
The topological order is a generic concept for such phases\cite{Wen89}
and the geometrical phases based on the Berry connection
supply explicit topological order parameters for them
\cite{Hatsugai0406}.

In such topological phases, many of them are gapped and form
a class of quantum liquids where
the  adiabatic invariants play fundamental roles
protected by quantization due to the finite gap.
They are  topological order parameters. 
Such successful examples are the Chern numbers 
for the integer QHE\cite{Thouless82} and $Z_2$-Berry phases for
the itinerant electrons and gapped quantum 
spins
\cite{Hatsugai0406}.
Topological insulators (in a restricted sense) for 
the quantum spin Hall effects as the time-reversal(TR) invariant 
QHE provide other important examples of the $Z_2$-quantization\cite{KM05,BHZ}.
Another fundamental feature of the 
topological phases is the bulk-edge correspondence\cite{Hatsugai93b}.
Even if the bulk is featureless, 
low energy excitations are localized near the boundaries or 
impurities, which are fundamentally governed by the non trivial bulk. 

One of these quantum liquids is a class of the spin liquids
for frustrated quantum spins where
geometrical frustration and/or quantum fluctuation
prevent the system from formation of conventional magnetic order. 
We have demonstrated the validity of the $Z_2$-Berry phases
where the TR symmetry as the anti-unitary 
invariance\cite{Hatsugai0406,MaruSpinBerry}
 protects 
their quantization.
As for the itinerant electrons with frustration, the same  consideration 
can be applied including spins and orbitals as additional local
degrees of freedom. 
Formation of diagonal order such 
as magnetism, charge order and  orbital order
can be suppressed due to the geometrical frustration and/or quantum
fluctuation. 
Then possible stabilization of the quantum states
is provided by a dimer formation and 
its generalization as a multimer formation.
To relax a local entropy,
quantum mixing  among the local degrees of freedom is quite
effective to
form a local quantum object. It is a multimer
in general. In other words,
it is a local covalent molecule (with spin). 
When the total system is composed of weakly interacting 
such local covalent multimers, 
it is natural to expect a gapped ground state. 
We do have such examples experimentally and 
theoretically\cite{Canals00,Tamura06,Motome06}

Although the covalent molecule as the multimer is simple, 
it is a purely quantum mechanical object such as the local spin singlet.  
Then it is natural to characterize the 
quantum phase by this multimer
as the quantum local order parameter. 
In cases of the spin liquids, 
the $Z_2$ Berry phases are quite successful 
to capture the local singlet
\cite{Hatsugai0406,MaruSpinBerry}. 
In this work, 
simple topological expression for this $Z_2$ Berry phases is explicitly given.
Moreover, corresponding to more general symmetries,
we provide  $Z_Q$ quantization 
of Berry phases, which 
are again quite successful 
to identify the $Q$-multimer.
We demonstrate the validity for 
a one-dimensional dimer as the model of 
Polyacetylene, fermions on the 
Kagome lattice in 2-dimensions and Pyrochlore lattices in 3-dimensions. 
The Kagome and Pyrochlore systems are 
canonical frustrated elect on systems and main targets
of many experimental and theoretical 
studies\cite{Canals00,Hiroi02,Atwood02,Udagawa10}.
Surprisingly these physically important systems 
are parts of the generic $Q=(d+1)$-multimer models in 
$d$-dimensions (hyper-Pyrochlore) and discussed in a unified way.

 {\em Hyper-Pyrochlore lattice in $d$-dimensions}: 
Let us provide a hyper-Pyrochlore lattice in 
$d$-dimensions. We consider fermions on this lattice
at the filing factor $1/Q$ per site,
 which has a $Q$-multimerization transition ($Q=d+1$)
as the quantum phase transition.
\begin{figure}
\begin{center}
\includegraphics[width=8cm]{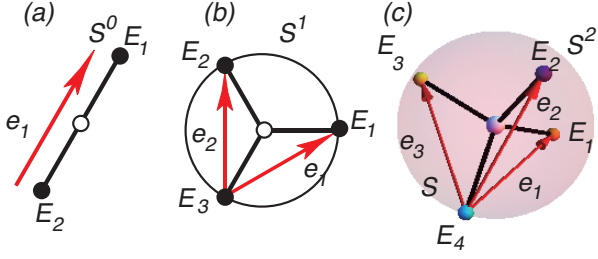}
\caption{
Vertexes of hyper-tetrahedrons $\{E_j\}$ 
 and unit translations $\{e_j\}$ 
 in $d$-dimensions. (a): $d=1$, 
(b): $d=2$ and (c): $d=3$.)
}
\label{fig:unit}
\end{center}
\vskip -1.0cm 
\end{figure}

Unit translations  of the system
are given as 
\begin{eqnarray*}
 {e}_j &=&  {E} _j- {E}_Q,\ j=1,\cdots, d
\end{eqnarray*} 
where $\{E_j, j=1,\cdots,d+1=Q\}$ is a set of equivalent $Q$ points
distributed on  the $d$-dimensional sphere $S^d=\{x\big||x|=1\}$
those form a hyper-tetrahedron in $d$-dimensions.
$S^0=\{\pm 1\}$ is a set of 2 points, $S^1$ is a unit circle and $S^2$ is
a unit sphere (Fig.\ref{fig:unit})
\footnote{
 $E_j^{(d)}$
are 
given as 
$E_{Q}^{(d)}= {^t}(0,\cdots,0,1_d,0,\cdots)$ 
and 
$E_j^{(d)} =  \alpha_d E_j^{(d-1)}+\beta_d E_{d+1}^{(d)}$,
($j=1,\cdots,d$), 
$
\alpha _d =\sqrt{(1-d ^{-1}  )(1+d ^{-1} )}
$
and 
$\beta_d=-d ^{-1} $.
\label{fnn:Ej}
}.
The set  $\{E_j\}$ 
are two points for $d=1$,
 vertexes of equilateral triangles in $d=2$ and 
those of regular tetrahedron in $d=3$.
Note that the unit vectors $\{e_j\}$ form unit translations 
of the generic Creutz hamiltonian in $d$-dimensions 
as the generic graphene\cite{GraCreutz}.

\begin{figure}
\begin{center}
\includegraphics[width=7.2cm]{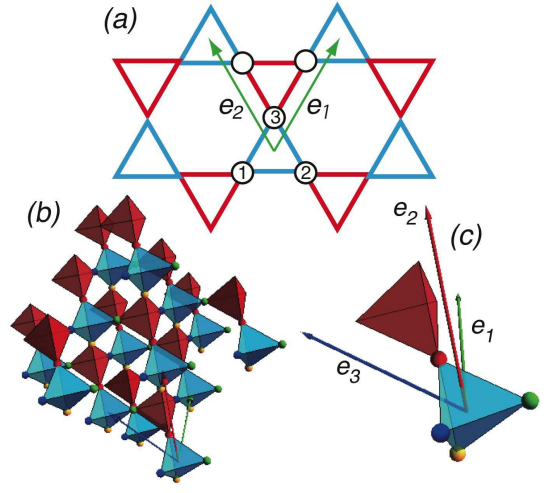}
\caption{Unit cells and unit vectors of hyper-Pyrochlore, (a) Kagome lattice
($d=2$), (b), (c) Pyrochlore lattice ($d=3$).}
\label{fig:genPyro}
\end{center}
\vskip -0.5cm 
\end{figure}

The hyper-Pyrochlore lattice is given by the decorating this 
Creutz model.
We put 
$Q=d+1$ intra sites at the positions $\{-e_j/2\}$ 
 ($j=1,\cdots,d, Q$) within the unit cell
where we define $e_Q=0$. 
This $Q$ intra sites  form  (blue) hyper-tetrahedron.
These blue hyper-tetrahedrons are connected by the
the other (red) hyper-tetrahedrons
which can be labeled at the same time
as $e_j/2$, ($j=1,\cdots,d, Q$)
( See Fig.\ref{fig:genPyro} ).
Here the hyper-tetrahedron in $d=1$ is a pair of sites,
that in $d=2$ is an equilateral triangle and it is tetrahedron 
in $d=3$. 

The hamiltonian of the fermions on the 
hyper-Pyrochlore lattice is given as
\begin{eqnarray*} 
 H  &=& \sum_{r}
 t_B[C_B(r)] ^\dagger C_B(r)
+
 t_R[C_R(r)] ^\dagger C_R(r),\  t_R, t_B\in\mathbb{R}
\end{eqnarray*}
where
$C_B (r)=  \sum_{j=1}^{d+1} c_j(r)$
and 
$C_B (r)= \sum_{j=1}^{d+1} c_j (r+e_j )$ with
$r=n_1 e_1+\cdots +n_d e_d$, ($n_j\in\mathbb{Z}$) 
which is a label of the unit cell.
The fermion annihilation operator at the site $r$ of the kind $j$
is denoted by $c_j(r)$.
The present hyper-Pyrochlore lattice can be understood 
as a line graph hamiltonian of the generic Creutz hamiltonian\cite{Katsura10}.
We consider a case, $t_{B,R}<0$, in this letter. 
Then the hamiltonian is negative semi-definite,
that is, one particle energy is at most zero
\footnote{  
The discussion 
by the Berry phase remains the same even when 
one includes particle-particle interaction unless the energy gap collapses. 
}.
In the  momentum representation,
$
c_j (r) = (N_1\cdots N_d)^{-1/2} \sum_k e^{i k\cdot r} c_j (k)
$, ($j=1,\cdots,Q$),
$k\cdot r=k_1n_1+\cdots+k_dn_d$ and 
$N_j$ is a number of unit cells in $j$-direction, 
we have
$
H = \sum _k c ^\dagger (k) h(k) c(k)
$, 
\begin{eqnarray*}
h &=& Q (t_B p_B+ t_R p_R)=\psi D \psi ^\dagger,\quad
D = Q 
{\small
\left(
\begin{array}{cc}
t_B & 0 \\
0 & t_R 
\end{array}
\right)
}
\end{eqnarray*} 
where $c ^\dagger (k)=(c_1 ^\dagger (k),\cdots,c_Q ^\dagger(k)) $, 
$\psi_B ^\dagger(k) =({1},{\cdots},{1})/\sqrt{Q}$,
$\psi_R ^\dagger  (k) = ({e^{ik_1}},{\cdots},{e^{ik_d}},{1})/\sqrt{Q}$,
and $\psi = (\psi_B,\psi_R)$.
The operators
$p_B = 
\psi_B\psi_B ^\dagger =p_B^2$,
$p_R=\psi_R\psi_R ^\dagger =p_R^2$
are projections into the 
linear space spanned by $\psi_B$ and $\psi_R$ respectively.
The states $\psi_B$ and $\psi_R$ are normalized, 
$\psi_B ^\dagger \psi_B =\psi_R ^\dagger \psi_R =1$,
 but generically  not orthogonal $\psi_B\psi_R\ne 0$,
\begin{eqnarray*}
\psi ^\dagger \psi &=& 
\left(
\begin{array}{cc}
1 & \Delta \\
\Delta ^* & 1
\end{array}
\right)
\equiv{\cal O},\quad
\Delta(k) =\psi_B ^\dagger \psi_R=1+\sum_{j=1}^d e^{-ik_j}.
\end{eqnarray*} 
Since 
a linear space
is invariant for linear operations,
the hamiltonian is diagonalized 
within the space spanned by $\psi_B$ and $\psi_R$.
Then it
 is now clear that 
the hamiltonian $h(k)$ has at most two non zero eigen values
and the others are zero. 
These non zero energy eigen states are 
spanned by $\psi_B$ and $\psi_R$
\footnote{ 
It is also confirmed explicitly as
$
\det (\lambda I_N-\psi D\psi ^\dagger )
= 
\lambda ^N\det\nolimits_N (I_N- \lambda ^{-1} \psi D\psi ^\dagger )
= 
\lambda ^N\det\nolimits_2 (I_2- \lambda ^{-1}  D\psi ^\dagger \psi )
= 
\lambda ^{N-2}\det\nolimits_2 (I_2 \lambda - 
{\cal O} ^{1/2} 
D
{\cal O} ^{1/2} 
)
$ where $N$ is a dimension of the Hilbert space.
When the multiplet is 
$M$-dimensions,
at least, $N-M$ zero modes exist similarly. 
}.
Then the spectrum is composed on the two non trivial bands and the $Q-2$ 
flat bands at zero energy.
In this work, we
consider a many body state of the filling factor $1/Q$, that is, 
the only the lowest band is completely filled.

When the states $\psi_B$ and $\psi_R$ are not linearly independent,
which only occurs at the origin $k=0$, 
one of the two bands touches to the zero mode flat bands
($k=0$ is the only touching momentum).
Except this touching momentum, the overlap matrix is regular as
$\det {\cal O} = 1-|\Delta |^2\ne 0$. 

Projecting out the zero modes, the hamiltonian for the
non zero energy bands is diagonalized 
by a linear combination of $\psi_B$ and $\psi_R$. 
Then the Schrodinger equation reduces to the following 
generic secular equation
$
{\cal O} D{\cal O}\phi=E{\cal O}\phi
$ where $\phi={^t}(\phi_B, \phi_R)$ is a two dimensional vector.
The two band energies are given by the eigen values of
$
h_\psi =  {\cal O}^{1/2} D {\cal O}^{1/2}
$
assuming  $\det{\cal O}\ne 0$.
Then noting that
$
\det h_\psi
= 
\det D\det {\cal O}=Q^2 t_Bt_R\det{\cal O}$
and 
$
\Tr  h_\psi = \Tr
 h = 
Q t_B \Tr p_B
+
Q t_R \Tr p_R
= Q(t_B+t_R)
$, 
the non trivial two band energies are given by
\begin{eqnarray*}
E(k) &=& ({Q}/{2}) \big( t_B+t_R\pm
\sqrt{(t_B  -t_R )^2+t_B t_R|\Delta (k)|^2}\big)
\end{eqnarray*}
which is the energy dispersion of the hyper-Pyrochlore system. 
When the two bands are degenerate, the Dirac fermions appear and the
many body state with the filling $1/Q$ state becomes critical,
which occurs only when $t_B=t_R$.
This is a quantum critical point of the $Q$-multimerization transition. 
We have two different $Q$-multimerization phases, which are 
 gapped ground states both  for $t_B < t_R$ and $t_B > t_R$.
In the following, 
we give topological order parameters for this 
quantum phase transition using $Z_Q$ adiabatic 
invariants as Berry phases.

Before describing the construction of the topological order parameter, 
let us discuss zero modes of generic multimer system in real space.
Consider a hamiltonian
$
H = \sum_\alpha t_\alpha C_\alpha ^\dagger C_\alpha
$,
$t_\alpha \in \mathbb{R}$,
where $
C_\alpha = 
(
\xi_1 c_{\alpha _1}+\cdots+\xi_{\alpha _{N_\alpha }}c_{\alpha _{N_\alpha }})
=\psi_\alpha c$ 
is a annihilation operator of an $N_\alpha $-multimer,
$
c ^\dagger  =({c_1}^\dagger ,\cdots )$
where $\alpha $ is a label of the generic $N_\alpha $-multimer in real space.
Then $\psi_\alpha $ is a normalized multiplet, which is the 
molecular orbital of the $N_\alpha $-multimer as
$
\psi_\alpha ^\dagger \psi_\alpha   =
|\xi_1|^2+\cdots+|\xi_{N_\alpha }|^2=1$. Note that the lattice structure is
arbitrary. 
Since the hamiltonian is again given by a sum of projections as
$
H = c ^\dagger \big[\sum_\alpha t_\alpha p_\alpha \big] c 
$
where
$ p_\alpha = \psi_\alpha \psi_\alpha ^\dagger 
$,
the discussion above is directly
applied\cite{Note3}.
 Then the number of zero modes, $N_Z$ is, at least, given by $N_Z^{\rm min}$
\begin{eqnarray*}
N_Z &\ge&  N_Z^{\rm min}\equiv \#(\text{sites})-
\#(\text{molecular orbitals})
\end{eqnarray*} 

{\em $Z_Q$ Berry phases}: 
Let us define a Berry phase introducing  a $d$-dimensional 
parameters 
$\Theta=(\theta_1,\cdots,\theta_d)$ where $\theta_j$ is defined on mod $2\pi$, 
that is, $\Theta$ is define on the torus $\Theta\in T^d$. 
It modifies a local blue hyper-tetrahedron  only at 
$r$ as
\begin{eqnarray*} 
C_B(r) &\to & C_B' (r)= \sum_{j=1}^{Q=d+1} e^{i\varphi_j}c_j(r) 
\end{eqnarray*} 
where
$
\varphi_j = \sum_{i=1}^j\theta_i
$, ($j=1,\cdots,d$) and 
$
\varphi_Q = 0
$.
If this hyper-tetrahedron is decoupled from the other ($t_R=0$), 
it is a gauge transformation. 
However, it is not the case
and the phases are not gauged away  generically. 
\begin{figure}
\begin{center}
\includegraphics[width=9.0cm]{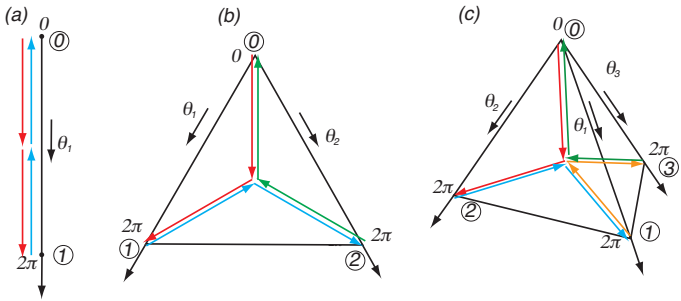}
\caption{
$Q=d+1$ paths 
 $L_1,\cdots,L_{Q}$ to define $Z_Q$ Berry phases in the parameter space $T^d$.
}
\label{fig:path}
\end{center}
\vskip -0.95cm 
\end{figure}
Using a loop (closed paths) $L_j$, ($j=1\cdots,d$) on the
torus $T^d$ defined
below,  Berry phases $\gamma_j $'s are defined as
$\gamma _j= -i \int_{L_j} A$, 
$A =\langle \Psi(\Theta) |d\Psi(\Theta) \rangle  $
where $|\Psi (\Theta)\rangle $ is a many body state which are energetically
well separated from the above for $^\forall \Theta$.
The loop $L_j$ is defined as
\begin{eqnarray*}
L_j &=& \ell_j-\ell_{j+1},\ (\ell_{Q+1}=\ell_1),\quad
\ell_j = V_j\to C_G
\end{eqnarray*} 
where $V_j=2\pi e_j$.
We use the same $e_j$ to save symbols. It
is in the parameter space $\Theta$ and not in the real space.
This $V_j$ 
is the $j$-th  vertex of the generic hyper-tetrahedron 
in the parameter space. 
The bond length of
the hyper-tetrahedron in the parameter space is
$2\pi$ and $C_G$ is a center of gravity of the hyper-tetrahedron, 
$C_G=\sum_{j=1}^Q V_j/Q$  (See Fig.\ref{fig:path}).
The hamiltonian on this loop, $H(L_j)$, is periodic.
The corresponding Berry phase is well defined
as far as the gap remains open. 

We note here symmetric properties of the hyper-Pyrochlore system. 
Single hyper-tetrahedron is invariant for any change of the 
vertexes $S_Q$
then the hyper-tetrahedron system has this global $S_Q$ 
symmetry, since the $S_Q$ induces a change of the 
bases $\{e_j\}\to \{\pm e_{j'}\}$.
  By the introduction of the phases $\varphi_j$, 
this symmetry is generically broken but we still has 
a following global $Z_Q$ equivalence defined 
as
\begin{eqnarray*} 
H(L_j) &=& U_QH(L_{j+1}) U_Q ^\dagger ,\quad (U_Q)^Q=1
\end{eqnarray*}  
where $U_Q$ is a parameter independent global unitary
operator.
It operates for the hyper-tetrahedron at $r$ as
$
U_Q c_{j+1}(r) U_Q ^\dagger = c_j(r)
$,
$ \ c_{Q+1}(r)\equiv c_1(r)$.
It implies the $Z_Q$ equivalent of the corresponding Berry phases as
\begin{eqnarray*} 
\gamma _1 &\equiv & \gamma _{2}\equiv \cdots \equiv \gamma _Q \equiv \gamma, \quad {\rm mod}\, 2\pi
\end{eqnarray*} 
where note that the Berry phase is generically gauge dependent and only
well defined in modulo $2\pi$
\cite{Hatsugai0406}.
Then noting that a sum of the loop 
is equal to the zero loop,
$\sum_{j=1}^Q L_j \equiv  0
$,
we have
$
\sum_{j=1}^Q \gamma _j \equiv Q \gamma \equiv  0
$, ($ {\rm mod}\, 2\pi)$. 
It implies $Z_Q$ quantization of the Berry phase as
\begin{eqnarray*}
\gamma &\equiv & 2\pi \frac {n}{Q} \quad {\rm mod}\, 2\pi, \quad n\in\mathbb{Z}
\end{eqnarray*} 

\begin{figure}
\begin{center}
\includegraphics[width=4.5cm]{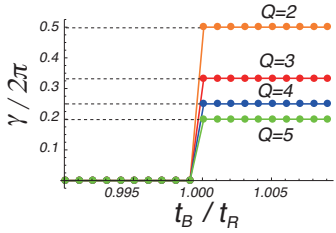}
\caption{Numerical calculation of the Berry phases 
for hyper-Pyrochlore lattices
$Q=2 (d=1)$: $N=2\cdot 200^1$, 
$Q=3 (d=2)$: $N=3\cdot 14^2$, 
$Q=4 (d=3)$: $N=4\cdot 6^3$, 
$Q=5 (d=4)$: $N=5\cdot 4^4$  
as a function of the multimerization strength $t_B/t_R$. The systems are periodic.
}
\label{fig:Q5}
\end{center}
\vskip -1.0cm 
\end{figure}

Since the quantization of the Berry phases is established, these Berry phases
are adiabatic invariants, that is, 
they never change unless the energy gap closes. 
When the hopping $t_B$ is sufficiently weak, 
the res ponce of the ground state against for the modification by $\theta_j$'s 
is  very weak,
that is, 
the Berry phases vanish due to the quantization, $\gamma \equiv 0$
(mod $2\pi$).
On the other hand, $t_B$ is strong enough compared with $t_R$,
 say, $t_R=0$, the Berry phases are determined by a single
hyper-tetrahedron at $r$. 
In this case, the change of the $\theta_j$'s is
given  by the gauge transformation of the hamiltonian 
as\cite{HiranoDeg}
$
H(L_j) = {\cal U}(\Theta) H(0) {\cal U}(\Theta) ^\dagger 
$,
${\cal U}(\Theta) = \prod_{j=1}^d e^{-i\varphi_j n_j(r)}
$,
$
|\Psi(\Theta) \rangle = {\cal U}(\theta)|\Psi(0) \rangle 
$,
$
A= -i \langle \Psi(0)|\sum_{j=1}^d n_j(r) | \Psi(0) \rangle
 d\varphi_j
=  - i\frac {1}{Q} \sum_{j=1}^d d\varphi_j
$
where $n_j(r)=c_j(r) ^\dagger c_j(r)$ is a occupation operator at the
site $j$ of the hyper-tetrahedron at $r$.
Note that the expectation values are 
$\langle \Psi(0)|n_j(r) |\Psi(0) \rangle =1/Q$ due to the $Z_Q$ symmetry
of the system without twist $\Theta$.
Then the Berry phases along the loop $L_j$ is given as
$
 \gamma = - 2\pi \frac {d}{Q} 
\equiv 2\pi \frac {1}{Q}$, (${\rm mod}\, 2\pi$)
in the strong coupling limit.
It is numerically confirmed as in the Fig.\ref{fig:Q5} for the
various hyper-Pyrochlore systems with periodic boundary conditions. 
The key idea of the $Q$-multimerization
 is that the $Z_Q$ equivalence of the
$Q$ loops in the hamiltonian manifold  and the sum of them is equivalent to 
a zero loop $\sum_j L_j= 0$. 
These two conditions are 
satisfied for hyper-Pyrochlore systems 
constructed with the parameter $\Theta$ in any dimension.

{\em $Z_2$-invariant for a Chiral symmetric system}:
Our hyper-Pyrochlore system for the $Q=2$ case
and the $d$-dimensional Creutz hamiltonians are
reduced into the  chiral symmetric system 
since the diagonal terms in the hamiltonian do not modify 
the one particle states at all.
Generically speaking, the $Z_2$-equivalence is derived 
only by the chiral symmetry\cite{Hatsugai10S}.
In this case, 
we have a direct and  explicit form of the adiabatic $Z_2$ invariant.
The chiral 
symmetry implies that the hamiltonian, 
$ H = c ^\dagger h c$ satisfies
$\{h, \Gamma \}=0$, $\Gamma ^2=I_N$.
Then  taking a form $\Gamma =
{\rm diag} (I_{N/2},-I_{N/2} ) $, the hamiltonian is written as
$h= \mmat{}{q}{q ^\dagger }{}$. 
Then the half filled 
ground state
$|\Psi \rangle$
is given by filling all of the negative energy states as 
$
|\Psi \rangle = 
(c ^\dagger \Psi_1)
\cdots
(c ^\dagger \Psi_N) |0 \rangle 
$,
$h\Psi_j=- \epsilon_{j}\Psi_j $,
 ($\epsilon_{j}<0$) since 
$\Gamma \psi_j$ is an eigen state with the energy $\epsilon_{j}>0 $.
We take the one particle states are orthonormalized as 
 $\Psi_i ^\dagger \Psi_j=\delta_{ij}  $.
Here we assume the half filled many body state is gapped. 
In this case, the Berry connection of the many body state 
is given as
$
A =  \langle \Psi | d\Psi \rangle =
\Tr a
$
where
$
a = \Psi ^\dagger d \Psi
$ is a non-Abelian Berry connection of dimension $N/2$
which is composed of the one particle states as
$
\Psi = (\Psi_1,\cdots, \Psi_{N/2})
$.
Writing this  normalized multiplet as
$\Psi=\left(
\begin{array}{c}
\psi_A  \\
\psi_B
\end{array}
\right)/\sqrt{2}
$, 
the orthogonality of the states among the different energies implies
$\Psi ^\dagger \Gamma \Psi =0$.
It reduces to
$\psi_A ^\dagger \psi_A=\psi_B ^\dagger \psi_B=I_{N/2}$
using the normalization of $\Psi$.
Then we also have
$\psi_A  \psi_A ^\dagger =\psi_B \psi_B ^\dagger =I_{N/2}$ since 
$\psi_{A,B}$ are $(N/2)$-dimensional square matrices. 
Using the Schrodinger equation,
$h\Psi= \Psi D$, 
we have 
$ \psi_B= q ^{-1} \psi_A D$.
By the  projection, 
$P=\Psi \Psi ^\dagger $ and
taking a multiplet $\Phi =  \begin{pmatrix}{I_{N/2}}\\{O}\end{pmatrix}$ 
(${\cal O}$ is a $N/2$-dimensional square zero matrix),
the gauge fixed normalized multiplet
is given as 
$
\Psi_\Phi =  P\Phi (\Phi ^\dagger P\Phi)  ^{-1/2} 
=  \begin{pmatrix}{I_{N/2}}\\{\xi }\end{pmatrix} /\sqrt{2}
$
where
$
\xi = q ^{-1}  g$ and 
$g=\psi_A D ^{-1} \psi_A ^\dagger=g ^\dagger $.
Then the gauge fixed
 Berry connection $a_\Phi$ is explicitly evaluated as
$
a_\Phi =   \Psi_\Phi ^\dagger d\Psi_\Phi=
\xi  ^\dagger d \xi /2
=
\xi  ^{-1}   d \xi /2
$. 
It gives the Berry phase
$\gamma $ for a closed loop $L$ as 
\begin{eqnarray*}
\gamma  &=& 
- \frac {i}{2}  \int_L  \Tr \xi  ^{-1} d \xi  
= 
- \frac {1}{2}
 \int_L   d\,  {\rm Arg} \det q
\end{eqnarray*}
since $g$ is hermite.
It gives an explicit topological expression of the $Z_2$-Berry phase
by the winding number,
which gives $\pi$ when $\det q$ 
goes around the origin odd number of times 
in the complex plane. 
Otherwise it is zero.


We thank discussions with  M. Arikawa.
The work is supported in part by Grants-in-Aid for Scientific 
Research, No.20340098 from JSPS (YH)
and No.22014002 (Novel States of Matter Induced by Frustration) 
on Priority Areas from MEXT (JAPAN).


%

\end{document}